\documentclass[aps,prl,twocolumn,showpacs,letterpaper]{revtex4}
\usepackage{graphicx}
\usepackage{latexsym}
\usepackage{amsmath}
\def\be{\begin{equation}}
\def\ee{\end{equation}}
\def\bea{\begin{eqnarray}}
\def\eea{\end{eqnarray}}

\usepackage{natbib}

\begin{document}
\title{Many-body perturbation theory using the density-functional concept:
 beyond the GW approximation}

\author{Fabien Bruneval,$^1$ Francesco Sottile,$^1$ Valerio Olevano,$^1$ 
Rodolfo Del Sole,$^2$ and Lucia Reining$^1$}
\affiliation{$^1$Laboratoire des Solides Irradi\'es, UMR 7642 CNRS/CEA,
   \'Ecole Polytechnique, 91128 Palaiseau, France \\
$^2$Istituto Nazionale per la Fisica della Materia e Dipartimento di Fisica
dell'Universit\`a di Roma ``Tor Vergata'', Via della Ricerca Scientifica, 00133
Roma, Italy}

\date{\today}

\begin{abstract}

We propose an alternative formulation of Many-Body Perturbation Theory that uses the
density-functional concept.
Instead of the usual four-point integral equation for the polarizability,
we obtain a two-point one, that leads to excellent optical absorption and energy
loss spectra. The corresponding three-point vertex function and
self-energy are then simply calculated via an integration, for any level of approximation.
Moreover, we show the direct impact of this formulation on the
time-dependent density-functional theory.
Numerical results for the band gap of bulk silicon and solid argon illustrate
corrections beyond the GW approximation for the self-energy.

\end{abstract}
\pacs{71.10.-w, 71.15.-m, 71.35.-y, 78.20.Bh}
\maketitle

The electronic structure of materials and its
response to an external perturbation are key quantities for the
interpretation of many experimental results or for the design of technological devices. 
In this context, {\it ab initio} electronic structure calculations have become a tool of choice.
One can already obtain useful information about the band structure thanks to the widely used
Kohn-Sham (KS) framework of the Density-Functional Theory (DFT) \cite{HKS};
for the response function, one can use an independent particle
Fermi's golden rule (which is equivalent to
the Random Phase Approximation (RPA)).
Beyond this, state-of-the-art 
calculations for solids are based on the Many-Body Perturbation Theory (MBPT). In that case, quasi-particle (QP)
band structure energies are obtained 
from the solution of an equation similar to the KS one, but with the KS exchange-correlation 
(xc) potential $v_{xc}$ replaced by the electron self-energy $\Sigma$, most often using Hedin's GW approximation \cite{H}. 
In this approximation, $\Sigma$ is the product of the one-particle Green's function $G$ and the screened Coulomb interaction $W$ calculated in the RPA. 
The resulting band structures, and in particular the band gap, are generally much closer to the measured ones than the KS 
results \cite{AJW2}.
In order to get improved response functions, the electron-hole interaction can then be included by
adding the so-called ``vertex corrections'' beyond the RPA, which is done in practice by
solving the  four-point Bethe-Salpeter equation (BSE) for the polarizability $P$; this leads in general to 
excellent absorption and electron energy-loss spectra \cite{RMP-NOI}.
In particular, one correctly describes the important excitonic effects.
When needed \cite{ferdy}, one could then use the vertex and the improved response function
to construct a new self-energy beyond the RPA GW approximation.

Unfortunately, calculations of vertex corrections are cumbersome
essentially because of the four-point (electron-hole scattering) nature of the BSE 
(see e.g. \cite{RMP-NOI} for the BSE, or \cite{andrea-life,allvertex} for the self-energy beyond GW).
Calculations of the response function have been limited to relatively simple systems.
Cancellation effects on quasi-particle energies
between the vertices in $P$ and $\Sigma$ have been discussed only for selected cases, especially for
the
homogeneous electron gas \cite{mahan2}, or for bulk silicon but using a vertex derived from a {\it local} approximation to
the self-energy, namely  the KS potential in the local
density approximation (LDA) \cite{DeReGo}.
It is therefore an important challenge to find an efficient way
to evaluate the effects of vertices arising from more realistic KS potentials,
or better, from non-local self-energies.

Alternatively, it is known that {\it in principle} one could obtain the polarizability
directly from a two-point equation: this is the case when one works in the framework of
time-dependent DFT (TDDFT) \cite{RGK}, since one propagates the density
instead of the Green's function. TDDFT could therefore  clearly be a prominent  alternative to
the BSE for the calculation of $P$.
Recently, a reliable (long-range) approximation for the two-point 
xc kernel $f_{xc}({\bf r},{\bf r}',t,t')$ of the TDDFT linear response equation for $P$
has been  derived  from the BSE; this combines the precision of the latter 
with the computational advantages of TDDFT \cite{allkernel}.
TDDFT is of course not designed to access one-QP properties, like the band structure;
yet, one may try to use the progress made 
concerning $P$ in order to find approximations for $\Sigma$
beyond the GW approximation, and a first attempt in this sense has already been
made concerning  QP lifetimes \cite{andrea-life}. 

The ultimate goal
would be of course to combine the density-functional and the QP
concepts in such a way that systematic and efficient
improvements to the spectroscopic quantities of interest
could be obtained. 
In this work, we show how this goal can be reached. 

We start from the Dyson equation 
\begin{equation}
\label{dyson12}
G^{-1}(12) = G^{-1}_0(12)-V(1)\delta(12)-\Sigma(12)
\end{equation}
where $(12)$ stands for two space, time and spin coordinates 
$({\bf r}_1t_1\sigma_1,{\bf r}_2t_2\sigma_2)$,
$G_0$ is the free-particle Green's function, and
$V(1)=U(1)+V_H(1)$, the total classical potential,
where 
$U$ is a time-dependent external potential that goes to the static physical potential at times 
$\pm \infty$ and whose fictious time-dependent part will be made vanishing at the end of the derivation,
and $V_H$ is the Hartree potential.
One can express the self-energy in terms of variations of the Green's function with respect to
the external potential, $\Sigma = -iv G\delta G^{-1}/\delta U$ \cite{H}, or
\begin{equation}
\label{eq:sigma}
\Sigma(12)= i G(14) 
\Gamma(42;5) \frac{\delta V(5)}{\delta U(3)}v(31^+),
\end{equation}
with the irreducible vertex function 
\be
\label{eq:gamma}
\Gamma(12;3) \! =\!  - \frac{\delta G^{-1}(12)}{\delta V(3)}
 = \delta(13)\delta(23)\!  +\!   \frac{\delta \Sigma(12)}{\delta V(3)}
\ee
and $v$ the bare Coulomb interaction 
(integration over indices not present on the 
left is implicit throughout the paper).
Disregarding $\Sigma$ on the right-hand side of Eq.\,(\ref{eq:gamma}) would yield the GW approximation.

The derivative $\delta \Sigma / \delta V$ is usually replaced by the chain rule
$(\delta \Sigma / \delta G)(\delta G/\delta V)$, which, using the relation
between the derivatives of $G$ and of $G^{-1}$, leads to the term $(\delta
\Sigma / \delta G)G G \Gamma$ that transforms Eq.\,(\ref{eq:gamma}) into an integral
equation for $\Gamma$ \cite{H}. This equation, or an equivalent form, with its four-point
kernel dominated by $\delta \Sigma(12) / \delta G(34) $, has to be solved in order to
get the irreducible polarizability $P=-iGG\Gamma$ and an improved
self-energy from Eq.\,(\ref{eq:sigma}). This is the main obstacle on the way to a
calculation of polarizabilities or self-energies beyond the RPA.

{\it The fundamental idea of the present Letter is to benefit from the Runge-Gross theorem 
of TDDFT \cite{RGK} in order to rewrite $\delta \Sigma / \delta V$ in Eq.\,(\ref{eq:gamma})}.
The one-to-one relation
between time-dependent densities and external potentials, or consequently between the 
densities and the classical potentials $V$, allows one to use an alternative
chain rule to express $\delta \Sigma / \delta V$, namely
$(\delta \Sigma/\delta \rho) (\delta \rho/\delta V)$
\footnote{
It should be stressed that the derivative of $\Sigma$ with respect to
$\rho$ in the density-functional sense is different from computing the
derivative of $\Sigma(12)$ with respect to $G(34)$ and then taking the limit $4 \to 3^+$.
Such a procedure would indeed lead to a rough approximation.}.
This transformation is hence exact whenever the linear response version of TDDFT, which is the only ingredient
needed here, is exact
\footnote{An important requirement of TDDFT is a well defined initial state.
This is not an issue here as the initial state is,
by definition of the equilibrium Green's function,
the $N$ particle ground-state in absence of any time-dependent perturbing potential $U$.
Moreover, the linear response TDDFT needs weaker requirements than the general case \cite{vleeuwen2}.
Limiting cases as the example of a stationary current in an infinite solid,
that one might not be able to handle in TDDFT, might instead require additional terms in the functional derivative;
this is however beyond the scope of this work. 
}.
It leaves observables like QP energies of the original equations accessible
and suggests straightforward approximations.
Equation\,(\ref{eq:gamma}) hence becomes
\begin{equation}
\label{eq:gamma-2}
\Gamma(12;3) 
=\delta(13)\delta(23) +  \frac{\delta \Sigma(12)}{\delta \rho(4)}P(43),
\end{equation}
where $P=\delta \rho / \delta V$ is the irreducible
polarizability that, as explained above, is usually calculated by solving the
vertex equation. However, by integrating Eq.\,(\ref{eq:gamma-2}) with two
Green's functions $G$, one directly obtains
\begin{equation}
\label{eq:P}
P(12)= P_0(12) + P_0(13)f_{xc}^{\text{eff}}(34)P(42),
\end{equation}
with $P_0(12)=-iG(12)G(21)$ and the two-point kernel
\begin{equation}
\label{eq:fxc}
f_{xc}^{\text{eff}}(34)=-iP_0^{-1}(36)G(65)G(5'6)\frac{\delta\Sigma(55')}{\delta \rho(4)}.
\end{equation}
In other words, {\it one can now first determine the two-point irreducible
polarizability $P$ from the integral Eq.\,(\ref{eq:P}), and subsequently the
three-point vertex $\Gamma$ via the integration of Eq.\,(\ref{eq:gamma-2})}. From $P$,
the reducible polarizability $P^{\text{red}}$
\footnote{The type of time-dependence of quantities appearing in the equations has to be specified;
a clean way is the formulation on the Keldysh contour \cite{vleeuwen2}.
In any case, one has of course
to use the adequate form when evaluating the formulas in practice.}
is obtained via $P^{\text{red}} = P + P v P^{\text{red}}$. 

Finally, the self-energy becomes
\begin{multline}
\label{eq:corr-nonloc}
\Sigma(12)
= iG(12)W^{\text{TC-TC}}(21) \\
+ iG(14) \frac{\delta\Sigma(42)}{\delta\rho(5)}P^{\text{red}}(53)v(31^+).
\end{multline}
The first term has the GW form, but with the testcharge-testcharge (TC-TC) screened 
Coulomb interaction $W^{\text{TC-TC}} = (1+vP^{\text{red}})v$, instead of the RPA one.
It has been discussed \cite{mahan2,DeReGo} that
$\Sigma = i G W^{\text{TC-TC}}$
would
yield unreliable results, because of the cancellation effects coming from the
second term.
In fact, the term $vP^{\text{red}}v$ contributing to $W^{\text{TC-TC}}$
creates the induced Hartree potential felt
by a classical charge. The additional term $( \delta \Sigma / \delta \rho) P^{\text{red}}$ is
responsible for the missing induced xc potentials that act on an electron or
hole.
It is therefore useful to reformulate Eq.\,(\ref{eq:gamma-2}) as
\begin{equation}
\label{eq:gamma-nonloc}
\Gamma(12;3) \! = \! \delta(13)\delta(23) 
\! + \!  \delta(12)f_{xc}^{\text{eff}}(14)P(43) \! + \! \Delta\Gamma(12;3)
\end{equation}
where
\begin{equation}
\label{eq:deltagamma}
\Delta\Gamma(12;3) = \left( \frac{\delta \Sigma(12)}{\delta \rho(4)}
   - \delta(12)f_{xc}^{\text{eff}}(14) \right) P(43).
\end{equation}
The most important effects are in fact contained in the
first two (one- and two-point) contributions to $\Gamma$ (called
$\Gamma^{(2)}$ in the following), whereas the three-point remainder $\Delta\Gamma$
can be interpreted as a subsequent ``non-locality'' correction.
$\Delta \Gamma$ has no effect on $P$,
as one can see by integrating
Eq.\,(\ref{eq:deltagamma}) with two Green's functions and using Eq.\,(\ref{eq:fxc}).
In the self-energy of Eq.\,(\ref{eq:corr-nonloc}), the inclusion of $\Gamma^{(2)}$ leads to
$\Sigma = i G \tilde W $ with a modified screened Coulomb interaction $\tilde W
= (1+(v+f_{xc}^{\text{eff}})P^{\text{red}})v$.
This is a
testcharge-testelectron (TC-TE) screened Coulomb interaction instead of $W^{\text{TC-TC}}$; 
this expresses the fact that an additional electron  or hole in the system cannot be
described as a classical charge. $\Delta\Gamma$ yields then in $\Sigma$ a correction term
to this physically intuitive contribution; it is entirely due to the
non-locality of the self-energy in Eq.\,(\ref{eq:deltagamma}).

Equation\,(\ref{eq:P}) is a two-point, but QP-derived, equation for the polarizability.
The link with TDDFT can be made by the fact that the diagonal of
$G$ yields the exact time-dependent density $-i G(11^+) = \rho (1)$ \cite{vLeeuwen}.
$\delta G / \delta \rho = - G (\delta G^{-1} / \delta \rho)  G$ leads to
\begin{equation}
\label{eq:tdss2}
 i G(13) G(41^+) \frac{\delta G^{-1}(34)}{\delta \rho(2)} =  \delta(12) .
\end{equation}

Since the same exact density, and hence the same Hartree potential, should also
be obtained from the Kohn-Sham potential $v_{KS} = V + v_{xc}$ we can write
\begin{equation}
G^{-1}(12) \! = \! G_0^{-1}(12)
  \!-\! \delta(12) ( v_{KS}(1)\! -\! v_{xc}(1) ) \!-\! \Sigma(12).
\end{equation}

As $\delta G_0^{-1} / \delta \rho = 0$, Eq.\,(\ref{eq:tdss2}) becomes:
\begin{multline}
\label{eq:tdss3}
 P_0(13) \chi_0^{-1}(32)
 - i G(13) G(41^+)  \frac{\delta \Sigma(34)}{\delta \rho(2)}   \\
 - P_0(13) f_{xc}(32)
 = \delta(12) ,
\end{multline}
where $\chi_0(12) = \delta \rho(1) / \delta v_{KS}(2)$
is the KS independent particle polarizability  and
$f_{xc}(12)= \delta v_{xc}(1) / \delta \rho(2) $ is the xc kernel of TDDFT.
This kernel turns out to consist of two terms, namely $f_{xc}^{(1)}$ and
$f_{xc}^{(2)}$, with $f_{xc}^{(2)}$ exactly equal to the $f^{\text{eff}}_{xc}$
arising from our previous approach and
\begin{equation}
\label{eq:f1a}
f_{xc}^{(1)}(12) =  \chi_0^{-1}(12)  -  P_0^{-1}(12).
\end{equation}

$f_{xc}^{(1)}$ serves to change the KS response function into the independent QP one,
in particular, to solve the so-called band gap problem.
$f_{xc}^{(2)}$ accounts for the electron-hole interaction.
This splitting \cite{pankratov} is physically intuitive. 
Altogether, TDDFT yields then  for the irreducible polarizability  P
\begin{equation}
P = \chi_0 +\chi_0 (\chi_0^{-1} - P_0^{-1} + f_{xc}^{\text{eff}}) P.
\end{equation}
{\it This is equivalent to Eq.\,(\ref{eq:P})}.

To get an explicit expression for $f_{xc}^{\text{eff}}$,
we choose
a starting approximation for the self-energy, and consistent approximations for the functional derivative of $\Sigma$ and for $G$, on the right side
of Eqs.\,(\ref{eq:fxc}) and\,(\ref{eq:corr-nonloc}). 
A simple choice could be to take  $\Sigma$, $G$ and $P_0$ as derived from a {\it local} 
and  {\it adiabatic} xc potential, e.g. the LDA one. This leads of course to the
TDLDA and the GW$\Gamma$ approach of Ref. \cite{DeReGo}. 
A better choice is to start from the GW approximation for $\Sigma$,
taking $W$ as a screened (e.g. static RPA) Coulomb
interaction. For the functional derivative, it is then reasonable
(i) 
to neglect the derivative of $W$ as usually done in the BSE:
(ii) to approximate $\delta G / \delta \rho = - G(\delta G^{-1} /
\delta \rho ) G$ by $G P_0^{-1} G$, truncating the 
chain of derivatives $\delta \Sigma / \delta \rho$ that would appear if one
continued to calculate all terms of $ \delta G^{-1} / \delta \rho $ and that
would lead to an integral equation similar to Fig. 2(b) of Ref.\,\cite{pankratov}.
(Note that this is equivalent to supposing $G$ be created by a local potential).
We obtain hence from Eq.\,(\ref{eq:fxc})
\begin{multline}
\label{eq:kernel}
f_{xc}^{\text{eff}} (34) = P_0^{-1}(36)G(65)G(5'6)W(55')  \\
   \times  G(57)G(75')P_0^{-1}(74) .
\end{multline}
Equation\,(\ref{eq:kernel}) is the electron-hole  xc kernel of
Refs.\,\cite{allkernel,pankratov}. In those works, $f_{xc}^{\text{eff}}$  was
derived in two completely different ways (by mapping matrix elements of the BSE
electron-hole interaction on those of TDDFT, and by a first-order expansion in
$W$, respectively) and extensively tested for 
the dielectric function, with excellent
results. In other words, the results of Refs.
\cite{allkernel,pankratov} yield the numerical validation of the
present derivation whereas, in turn, the latter {\it a posteriori} yields new
physical insight about {\it why} the former had led to such (unexpectedly)
good results: {\it the physics of the variation of the self-energy
upon excitation, which gives rise to the electron-hole interaction, 
can be captured in terms of density variations only}. This is very important,
since it encourages the use of the present scheme also for cases where one wishes
to go beyond the above approximations.

It is interesting to see what one
obtains when using this scheme to go beyond the GW approximation for the
calculation of band gaps in semiconductors and insulators. Since
systematic GW$\Gamma$ studies in literature are available
only for a short-range (LDA) kernel \cite{DeReGo}, we provide here a  discussion
on the influence of a {\it long-range} contribution on QP energies.

\begin{table}
\begin{tabular}{lccccc}
\hline\hline
   & \phantom{C}LDA\phantom{EX}  & $GW^{RPA}$ & $GW^{\text{TC-TC}}$ & $G\tilde W$ & Expt. \\
\hline
Si & 2.53 & 3.17                   &        3.08                     & 3.18                   &  3.40 \\
Ar & 8.18 & 12.95                  &       12.64                     & 12.75                  & 14.2  \\
\hline\hline
\end{tabular}
\caption{\label{tab1}
Direct gap (in eV) at $\Gamma$ in bulk silicon and solid argon, calculated
using a local approximation (LDA) for the starting self-energy (see text).}
\end{table}

\begin{table}
\begin{tabular}{lccccc}
\hline\hline
   & COHSEX  & $GW^{RPA}$ & $GW^{\text{TC-TC}}$ & $G\tilde W$ & Expt. \\
\hline
Si & 3.64 & 3.30                   &         3.18                    & 3.32                   &  3.40 \\
Ar & 14.85& 14.00                  &        14.16                    & 14.76                  & 14.2  \\
\hline\hline
\end{tabular}
\caption{\label{tab2}
Same as Table\,\ref{tab1}, but based on a non-local
approximation (COHSEX) for the starting self-energy.
}
\end{table}

For illustration, we present in Tables\,\ref{tab1} and\,\ref{tab2} results on bulk silicon and solid argon
(obviously, the effect of a long-range contribution is particularly interesting
in a solid, and silicon and argon represent two extreme cases, the first one
with strong screening and continuum excitons, the second one with almost no
screening and strongly bound electron-hole pairs).
 
The first series of results, presented  in Table\,\ref{tab1}, uses LDA as starting approximation for the right-hand side of Eq.\,(\ref{eq:corr-nonloc}),
whereas the second series in Table\,\ref{tab2} uses the static but nonlocal
``Coulomb-hole-plus-screened exchange'' (COHSEX) approximation to GW \cite{H}.
In the latter case, we use the kernel $f_{xc}^{\text{eff}}$ given by Eq.\,(\ref{eq:kernel}) which, although approximate, has the correct long-range behaviour~\cite{allkernel,GGG}.
Furthermore, LDA wavefunctions are used throughout: we suppose them to be similar to the COHSEX QP ones
\footnote{We do not enter here the debated subject of fully self-consistent QP calculations
(see, e.g., \cite{HolmBarth})}.

The two tables show the band gap at $\Gamma$ for both materials under study. 
The first column gives the band gap that is obtained from the respective starting approximation (i.e. LDA or COHSEX). The second column uses this band structure, and provides the subsequent standard non-self-consistent $GW^{RPA}$. 
Columns 3 and 4 show the band gap for the approximations to the self-energy
derived in this work, using either $W^{\text{TC-TC}}$ (first part of Eq.\,(\ref{eq:corr-nonloc}))
or $\tilde W$ (neglect of only $\Delta\Gamma$).
Finally, the experimental value is given in the last column \cite{exp}.
Both materials show similar tendencies. In particular, there is a significant influence
of the single-particle energies on the $GW^{RPA}$ (second columns).
The choice of COHSEX energies in $W$ simulates the effect of the contribution $f_{xc}^{(1)}$ of Eq.\,(\ref{eq:f1a}).
In most cases, the electron-hole vertex correction $f_{xc}^{\text{eff}}=f_{xc}^{(2)}$ in $W^{TC-TC}$ closes
the gap (third columns) with respect to RPA. When $f_{xc}^{\text{eff}}$ is included
according to Eq.\,(\ref{eq:corr-nonloc}) in order to evaluate the explicit vertex in $\Sigma =iGW^{TC-TC}\Gamma^{(2)}=iG\tilde W$
(fourth columns), there is a strong opening of the gap.
Our most complete result is hence
determined by a series of cancellations.
The overall behaviour of both kernels under study (arising from LDA or the non-local COHSEX scheme)
is very similar, even though the LDA kernel does not have the crucial,
correct long-range contribution~\cite{GGG}.
These results roughly justify calculations using the RPA GW form constructed
with QP energies instead of KS ones.
The $GW\Gamma$ gap turns out to be slightly bigger than the
experimental value. In order to obtain improved agreement, one should of course
avoid some of the above approximations; in particular we expect the
non-locality correction to decrease the gap, since the neglected term
should reduce the effect of the external vertex. Those and other more
sophisticated numerical calculations (including, e.g., self-consistency in the
wavefunctions) are however beyond the
scope of this illustration.

In conclusion, using the concept of the density as crucial quantity we have
derived a complete new set of equations for the many-body vertex, polarizability
and self-energy. This approach does
not require the solution of integral equations containing a four-point kernel.
In particular, the polarizability is directly obtained from a two-point
equation, containing a two-point many-body kernel $f_{xc}^{\text{eff}}$, which completely changes the way e.g. excitonic effects can be calculated.
We have shown that the same expression for the polarizability can also be
derived from the relation between the Green's function and the charge
density.
Moreover the latter derivation  yields the
exchange-correlation kernel of TDDFT, which turns out to differ from
$f_{xc}^{\text{eff}}$ by a term that is essentially responsible for the gap
correction.  Our approach  explains the success of previously published
approximations for the kernel and
allows one to go beyond in a systematic way. On the other hand, it opens the way for better approximations to the self-energy and other many-body quantities.
For the gap
corrections in bulk silicon and solid argon, we have put into evidence
cancellation effects of different contributions to the vertex
corrections beyond the GW approximation.

We are grateful for discussions with
C.-O.~Almbladh,
U.~von Barth,
J.~F.~Dobson,
A.~Marini,
A.~Rubio,
G.~Stefanucci,
R.~van~Leeuwen,
and N.~Vast, 
support from the
the EU's 6th Framework Programme through the NANOQUANTA Network of Excellence (NMP4-CT-2004-500198),
and computer time from IDRIS (project 544).

\bibliographystyle{apsrev}

\end{document}